# A Framework for Debugging Java Programs in a Bytecode

Safeeullah Soomro
*Department of Computer Studies*
*AMA International University*
Salmabad, Bahrain
s.soomro@amaiu.edu.bh

Mohammad Riyaz Belgaum
*Department of Computer Studies*
*AMA International University*
Salmabad, Bahrain
bmdriyaz@amaiu.edu.bh

Zainab Alansari
*Department of Computer Studies*
*AMA International University*
Salmabad, Bahrain
zeinab@amaiu.edu.bh

Mahdi H. Miraz
*Centre for Financial Regulations and
Economic Development (CFRED)*
*The Chinese University of Hong Kong*
Sha Tin, Hong Kong
m.miraz@cuhk.edu.hk

*Abstract*— In the domain of Software Engineering, program analysis and understanding has been considered to be a very challenging task since decade, as it demands dedicated time and efforts. The analysis of source code may occasionally be comparatively easier due to its static nature, however, the back-end code (Bytecode), especially in terms of Java programming, is complicated to be analysed. In this paper, we present a methodological approach towards understanding the Bytecode of Java programs. We put forward a framework for the debugging process of Java Bytecode. Furthermore, we discuss the debugging process of Bytecode understanding from simple to multiple statements with regards to data flow analysis. Finally, we present a comparative analysis of Bytecode along with the simulation of the proposed framework for the debugging process.

## I. INTRODUCTION

Software Maintenance, as part of software development process, is an expensive but cost-effective procedure. In the software development industry, this process is often carried out manually due to unavailability of software testing and verification tools. Even software test data is difficult to find from other resources for research purposes both in the academia and in the industry. Therefore, automated tools for the software testing and verification may play a vital role both in the industry and the academia. These tools may reduce the cost of the proposal and provide accuracy within the time limit of the proposal. Automation of such tools, is likely to reduce software maintenance cost, which is very high at this moment. Furthermore, Software Verification and Testing tools may help software developers to reduce the time of the proposals, which is one of the most critical phases of the software development process in software engineering.

Currently, most of the Java applications are available in shape of Class files (like .jar and zip), however, this leads to a problematic situation for the user due to unavailability of the source code of the programs. While running such programs, if it any bugs is observed, it is highly necessary to fixe and overcome them, which is not a simple process. The user has to contact and report a bug to the tester and/or the developer, which is a very costly as well as lengthy process to follow. Furthermore, testers and developers are not always available to fix the problems. In this scenario, our framework to understand Java-Bytecode for testing, maintaining and debugging can help programmers, managers and maintainers to improve knowledge of understanding of the programs and help to overcome bugs.

Understanding of Bytecode is extremely important for the programmers developing automated tools needed for the software maintenance, testing and verification purposes. Such automated tools are dire necessities for multifaceted software engineering tasks such as program slicing, debugging, testing, maintenance and complexity measurement [1, 2, 3, and 4]. The work of [4] is closely related to this work as the authors provided path execution of Java programs in terms of Bytecode analysis. In fact, Bytecode is a middle way representation of Java programs. The Bytecode is shared with the backend knowledge of the source code, understanding of which may help the software engineering community in future. This information is crucial when we perform software engineering tasks such as debugging, maintenance, testing and verification. The Bytecode is compiler instruction from the Java files and fetched into virtual memory. If needed, it can be retrieved from the class file of the source program using any tool that can convert Java source code into Bytecode.

This paper presents a systematic approach of understanding Bytecode of each Java statements of the source code. We have presented the Bytecode of each statement and analyse it according to the execution of most of the Java language statements presented in Bytecode. Further, we explained the Bytecode of all the statements to increase readability as well as comprehensibility. Understanding of Bytecode is likely to help development of automated tools for debugging, testing and verification, which is a contemporary demand.

This article advances Java Bytecode awareness from the software engineering aspect. The developers and programmers of any programming languages requires understanding the middle-level language to further develop their capabilities in terms of execution and runtime analysis of programming languages. This is more pertinent in the aspect of compiler understanding tasks, which may help generating the fastest and smallest code while learning backend code (Bytecode).

The rest of the paper is organised as follows: section 2 briefly introduces Java Virtual Machine (JVM) and Bytecode while section 3 considers Bytecode statements of Java programs; section 4 discusses the possible applications while section 5 considers comparative analysis and finally section 6 contains concluding discussions as well as the final remarks.

## II. BYTECODE AND JAVA VIRTUAL MACHINE (JVM)

Bytecode is defined as an intermediate code of the java source code, which comprises of portable code (p-code) and intermediate code. The source.java file contains the front-end code of the Java while the .class file conserves the compact representation of the source code. The Bytecode statements of each instruction of Java code is portable binary codes, which are extracted from the .class file. The authors of [5] presented the variables, classes, methods and other related information of each Java statement in Bytecode information. This information can be used for testing and verification of faults from Java programs [6, 7] and for execution of code.

In another study [8], simulators were implemented for Java Bytecode which can specify values to detect faults from programs using backwards symbolic analysis. [8] reports the results of the aforementioned research, which is found to be appropriate for finding and locating faults from Java Bytecode.

A Java Virtual Machine (JVM) [9] is a virtual machine that compiles and executes the Java Programs as well as programs written in selective other languages into Java Bytecode. The JVM has two primary functions [9]: 1) to run programs in different free operating systems or device environments and 2) to manage the memory optimisation.

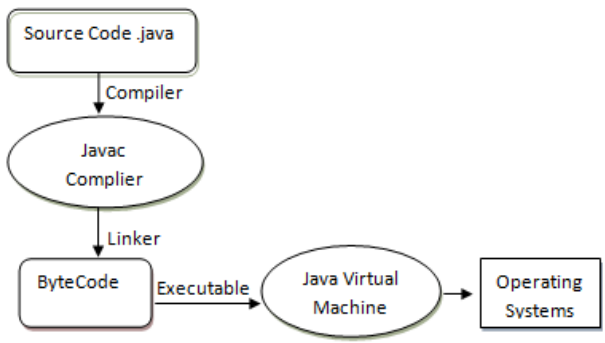

Fig. 1 Representation of Java Virtual Machine [6]

Fig. 1 represents the comprehensive compilation and execution of Java Programs using JVM on a different platform of operating systems. [8] discusses and reports detailed further information on JVM.

## III. BYTECODE STATEMENTS OF JAVA PROGRAMS

The debugging task, part of the software maintenance process, is considered to be a crucial phase in the software development, which is pertinent to reduce the time as well as cost of software maintenance. Contemporary research is focusing on developing automated tools for the debugging process, to enhance and expedite software testing and verification. Our effort is in-line with this avenue of the software maintenance research community focusing on the development of optimised framework and methodology for efficient automated tools for software maintenance.

In this regard, we define the typical software testing framework in terms of the input and the output of the programs in Bytecode. We used Bytecode due to the unavailability of source code. The aim of this research is to help locate the bugs from Bytecode if something goes wrong in the input/output of a program.

Fig. 2 demonstrates comprehension of data/values as well as dependencies of the input and the output of a simple program. The statements of the program contain the variable dependencies, which may affect the outcome of the simple multiple statements. For example

*Line 1 a= b;*          (a,b)

*Line 2 b=c,*          (b,c) (a,c)

The above simple statement shows the representation of variable dependencies. Line 1 represents the simple statement that *a* depends on *b*. In Line 2, *b* depends on *c*. so if the value added in *a* and *b* then diagonally it depends on *a* on *c* so that *(a, c)* comes through the dependencies of both lines. Authors of [8] provided information of many input cases and a small number of output cases, which is a part of forwarding and backward tracking, to find bugs through simulation. However, this still needs a framework for all the statements of the Bytecode.

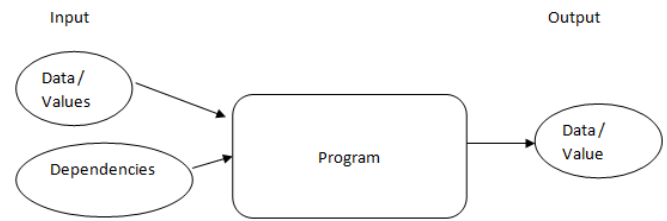

Fig. 2 Input/output Program Understanding

Fig. 3 provides the Bytecode information of the source code, which represents the statements of input and output in terms of Bytecode. We calculate the program specification on the overall basis of Bytecode, but each fragment has its own specification also known as "block specification". We assume that such program specification, which we can take through statements' specification, ultimately helps to find and fix the faults from Bytecode.

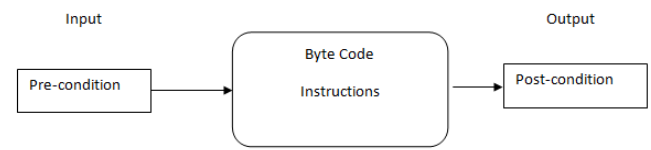

Fig. 3 Block/Program Specification of Bytecode

In the below Java Program, we provide an example to detect and locate the bugs in terms of value making use of our framework.

```
Specification in terms of Value is 3.
Specification in terms of dependencies are  (n1,n2),
(O1,n1), (O2,n2)
public static void main(String[] args) {
float n1=2; // instead of n1=4;
float  n2=3;
if (n1 >= n2) {
                        return (n1);
            } else {
                        return (n2);
            }
      }
```

Fig. 4 a small Java example

```
//Simple statement
    0 fload_0;
      1 fload_1;
     2 fcmpl;
/
If statement
    3 iflt 5;
    // True if
    6 fload_0;
    7 freturn;
    // False if
    8 fload_1;
  9 freturn;
```

Fig. 5 Bytecode of Java Program Fig. 3

In Fig. 4, we provided an example of a Java program having only one condition, which if executed would provide true or false depending on the value. According to variable dependencies, the program dependencies are n1, n2 and the results came through *if statement*.

According to our provided framework, we assume that if the values of n1=2 and n2=3 then output should result 3. The expected output of the simple fragment in fig. 4 is counted as 3. Simultaneously the specification of the program in terms of dependencies are *(n1, n2) (O1, n1) (O2, n2)*. This is the exact program specifications in terms of values and depending on the correct program. If the value is changed in the line number 2, then the output of the program is different and the program is counted as logically wrong. So we consider to take all the lines, one by one, through assumptions and keep each value on our knowledge through the framework and compare with the given program specification. Thus, we may find that the fault is a source code of the Java programs in Figure 3.

In fig. 4, Bytecode is provided with the corrected program, which represents the code conversation for each statement of the source code. Whenever any fault is observed into source code, we can see the conversation, which seems to be incorrect according to the specification of the Bytecode. However, our framework is limited in this case, especially on how to extract the specification from Bytecode and what to present in the specification regarding values and dependencies. The work presented in [8] provided the mechanism for finding bugs backwards and forward simulation of programs in a Bytecode.

In order to collect information from Bytecode, we have extracted values and dependencies from it. In Fig. 3, we have presented two ways of specification: one is value dependent and another is variable dependent. We have shown examples in source code and Bytecode. In the source code, we have shown how we extract values for input and output for each fragment. We have also shown how the values are executed and output generated is 3, which is equal to value specification in a corrected program. Whenever we introduce a bug in line number *1* and calculate the value of the small fragment, we have received value to be 2 which is not equal to our given specification of the small code of the program. Here, we find inconsistencies in between given specification and this code to be faulty. This is a small practice of the source code, which we have presented for understanding the analysis and debugging process. Another analysis of the source code is to find inconsistencies of variable dependencies in between given and computer dependencies. We have calculated specification *(n1, n2) (O1, n1) (O2, n2)* which is not odd given, so we have seen the inconsistencies between both specifications. In regard to inconsistencies, we have found the program is faulty. Furthermore, we have investigated the dependencies through assumption utilising our framework and found that either line number 4 or line number 5 maybe faulty in the source code. This methodology provides the possible location of faulty lines, detected through our proposed model and this information is further utilised to remove it. At this stage, we are not providing the solutions to fixing the faults of the program; our focus is to provide with the missing value and dependencies from the code, which may affect the program and find it faulty. Thus, we have accomplished it using a small example in the source code.

Fig. 5 demonstrates a Bytecode to find faults from the program we have provided in the pseudocode below (in Fig. 6) for finding faults from the program and notifying the misbehaviour in a Bytecode that is a cause of the fault. We have pointed out the missing value or variable in a Bytecode, which made our program faulty. Thus, using this approach, we have produced the faulty lines from the Bytecode information.

In Figure 6. *Fload* represents the float declaration in a program. The *fcmpl* shows the comparison in a Bytecode which is *if* statement in source code and *freturn* represents the control exit of the *if* statement whether the condition is true or false. The process of debugging in Bytecode information is to calculate the values from each line and keep in a memory table so that it can be used for the comparison with the original values of given values in a specification. After that, we have processed the simple statement from the declaration part the kept value in a table. Furthermore, we have calculated the values from each line in nested statements, repetitive statements and conditional statements of the program. We assume that we have calculated outcomes of all the lines in a term of value and save in a table. Now the table is filled all values taking from each line till the end of the Bytecode of the source code. Finally, we calculate and match the given values from our specifications with our calculated table through our framework. We have to assume each line in the Bytecode and find inconsistencies in between given and calculated specification. Furthermore, we have analysed all the lines, calculated the values and complete the process of comparing the given specification. If any consistency is found, the program is faulty. This information may help to pinpoint the broken lines in the Bytecode, which could be recovered through our findings. Due to this process, we have calculated the value based information from the Bytecode for the debugging process, to develop possible application tool in the future.

---

*Start*

*Calculate the line by line values*

*Record it in the table*

*Take the output of each line*

*Record it in the table*

*While repeat until find all tokens of the values from each line*

*Whenever found values match with Table for each line*

*Check whether equal to Specification of each line*

*If equal then that line is ok*

*Otherwise, that line may cause of faults in a Bytecode.*

*Repeat this procedure until the end of all lines.*

*End*

---

Fig. 6 Pseudo code of debugging framework in a Bytecode

Finally, we provide the exact information of the faults through the framework in terms of values from Bytecode segments: simple, multiple and nested statements.

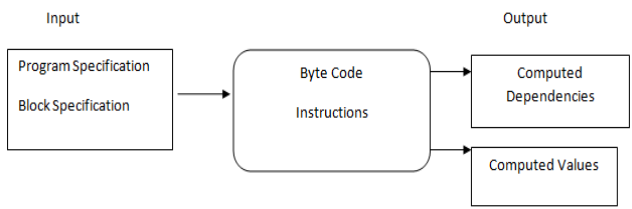

Fig. 7 The pseudo code of debugging framework in a Bytecode

In Fig. 7, we have provided a general framework of the proposed work to extract Bytecode information from the source code in terms of values and dependencies. The program specifications are generally calculated for the source code and particular block specification is added into the general specifications. After that, we have extracted the Bytecode information from the source code in terms of values and dependencies ensuring all information is extracted from the source. If source code is not provided then Bytecode information is dependent on the input and the out values. Finally, we compared the program specification and blocked specifications in terms of Bytecode to compare with input values and dependencies. The comparison is depending on the values of variable dependencies of the Bytecode and locating the mismatch in the given and calculated specification. If we find any discrepancies in the Bytecode, then we can assure that this program is faulty. Furthermore, we have used our framework to match the values and variables dependencies in the Bytecode to find the real misbehaviour of the Bytecode, which seems to incorrect input values of dependencies. In that process, if we find missing values and dependencies it can be confirmed that the program is faulty. Thus, we provided the mechanism of the debugging process mode of the framework for the Bytecode of Java programs.

## IV. POSSIBLE APPLICATIONS OF BYTECODE

Currently there is one tool [10, 11, 12] which can provide graphs from bytecode but failed to provide more than one class used in Java source code. [4] provided information on java Bytecode in the control flow graph for the understanding of bytecode only. The authors of [11] provided transformation of Java Bytecode into basic blocks and represented that information into Boolean functions through binary decision diagrams. There are some tools on Java Source code for analysis and debugging [7], but there is no such tool, which can provide better understanding of Java bytecode in terms of analysis and maintenance of the programs.

The work presented in [4] is a better platform to analyse Java Bytecode, which can understand the low-level code and map together with source code for the better understanding of Java Programs. Our framework can help to debug research community in the future for finding faults through our built-in the tool without having the source code.

Various tools to analyse the programs have been discussed in [14] for detecting the vulnerabilities. In fact, there are static tools as well as dynamic tools to serve this purpose such as Python Taint and WALA. The application of WALA can statically analyse the programs with a data flow graph for pointer analysis. In WALA, construction of the call graph is carried out to know the program behaviour. Machine learning being an emerging field also adopts this analysis. Python is a popular language for machine learning, however, there is not enough support from Python for error detection in it. So the application of WALA, a static tool has been used for tracking sensors in Python.

## V. COMPARATIVE ANALYSIS

The work presented in [13] provides with customisable tools and analyses the methods of Bytecode to obtain dynamic information of programs. Furthermore, it represents the Bytecode information in a visualisation form. However, this is limited in scope, which only provides visualization of small java programs. The advantage of our work is that we provide a framework to collect the program specifications from the Bytecode information for the simple, multiple and nested statements. The framework provided is to find faults from Bytecode in terms of values and dependencies, which is a novel approach.

In contrast, we used Java Bytecode for a better understanding of programs towards finding faults. We assume to present a visualisation of all statements of Java programming language like method calls, object creation, calling object, parameters passing through methods and objects, polymorphism and others in the object-oriented program. Out visual presentation is for the primary lines of code, multiple lines, loops, nested structures of code and fully support the aspects of object-oriented programs.

In the process of analysing the Java Bytecode, the authors of [15] have explained the issues aroused while performing the intraprocedural control flow analysis and interprocedural control flow analysis. In the process of analysis, rules have been framed to determine the basic blocks, which is a point to construct the control flow graphs. Exception handling situations, which may arise during the execution of Bytecode were also discussed. The applications of control flow graphs in different stages of software development with a focus on testing and maintenance was discussed.

Java Bytecode sequences are analysed using Bigram [16]. The authors of [16] considered the most commonly used pair of Bytecode to check the frequency of occurrences by performing benchmarking two widely used suites.

A framework called SOOT has been discussed in [17] by the authors who are used to perform analysis of programs as well as to experiment with the software engineering techniques. A detail explanation with pointer analysis and its side effects, along with the alternatives available to bypass the side effects, have been discussed.

While analysing the Bytecode, at some stage, it should also terminate. There are various types of programming paradigms like sequential, conditional, iterative etc. An automatic termination analysis is presented for the sequential Bytecode programs in [18]. Because of the language being used in developing mobile applications, which is in frequent use now a day, they need to be tested very well. A Rule-Based Representation (RBR) was introduced in contrast to the control flow graphs. A set of procedures with rules for each of the procedure is used to conduct termination analysis based on the constraints like input-out size relations, direct or indirect calls to the pairs.

In order to detect the bugs, formal method is also used in software testing. Forward simulation is an easy way of testing by giving the inputs and measuring the outputs. In [19], a symbolic backward simulation has been proposed. During the process of backward simulation, undefined variables are formed and those are represented as symbolic values. These symbolic values are determined in branch processing, loop processing and are proved to be useful in detecting the bugs.

The authors of [20] presented a methodology by interpreting the Java Bytecode during the dynamic execution. Their objective was optimising the test data during the searching process. Using the control flow graphs, the metrics based on the distance for method call, a control node, the local problem node are evaluated. Moreover, the test criteria are used for dynamic analysis of the test object's class file. Optimizing as an aim, the tool is used to test at the case level.

The authors[12,13,14] used the models for the bytecode information which may help researchers to retrieve back-end code with visualization. That means clear understanding of bytecode in a visual forms, which seems to be very promising work in the dynamic analysis of bytecode. Furthermore[12], this work shows the graphical user interface framework of the bytecode which helps researchers to retrieve information from the back-end code. Moreover, it helps to collect information of the data and variable flow graphs of the bytecode. The work[4] presented here which os very close to it as researchers of that paper is presented the control flow graph of the bytecode which may help users to carry software maintenance tasks like debugging, testing and verification.

## VI. CONCLUSION

In this work, we provided the framework of the debugging process of Bytecode for the Java programs in terms of values and variable dependencies. It is indeed a huge challenge to provide a debugging model in a dynamic analysis; therefore, we have advocated the framework of this process. We have proceeded with the pseudocode and executed it on the Bytecode for the finding or locating the faults from the Bytecode. We have presented a small example, which ensures that our methodology certainly provides the misbehaviour of the program in terms of Bytecode. Furthermore, we have provided the comparative analysis to ensure the visibility of this research work. Moreover, we have demonstrated possible applications of the research presented in this paper, which adds value to the software maintenance research domain and help advance the field.

In the future, we aim to develop an automated tool, which will extract information from the Bytecode in terms of value and variable dependencies in the aspect of the debugging process. While developing such tool, we aim to integrate support for the pure object-oriented programs, which obviously will be a significant challenge in terms of dynamic analysis.

ACKNOWLEDGEMENT

This work is supported by CTRG Research Group of the College of Computer Studies, AMA International University, Kingdom of Bahrain and Centre for Financial Regulation and Economic Development (CFRED) The Chinese University of Hong Kong Sha Tin, Hong Kong SAR, China.